\documentclass{article}

\usepackage{arxiv}

\usepackage[utf8]{inputenc} 
\usepackage[T1]{fontenc}    
\usepackage{hyperref}       
\usepackage{url}            
\usepackage{booktabs}       
\usepackage{amsfonts}       
\usepackage{nicefrac}       
\usepackage{microtype}      
\usepackage{lipsum}

\usepackage{xcolor}         
\usepackage{graphicx}       
\usepackage{subfigure}		
\usepackage{makecell}       
\usepackage{amsmath}        
\usepackage{xurl}           

\title{Forecasting Loss of Signal in Optical Networks with Machine Learning}

\author{%
	Wenjie Du \thanks{Work was done during a research internship at Ciena.} \\
	Concordia University \\
	Montr\'eal, Canada \\
	\texttt{wenjie.du@mail.concordia.ca} \\
	\And
	David C\^ot\'e \thanks{Corresponding author.} \\
	Ciena Corporation \\
	Ottawa, Canada \\
	\texttt{dcote@ciena.com} \\
	\And
	Chris Barber \\
	Ciena Corporation \\
	Ottawa, Canada \\
	\texttt{cbarber@ciena.com} \\
	\And
	Yan Liu \\
	Concordia University \\
	Montr\'eal, Canada \\
	\texttt{yan.liu@concordia.ca} \\
}

\begin{document}
\maketitle

\begin{abstract}
Loss of Signal (LOS) represents a significant cost for operators of optical networks. By studying large sets of real-world Performance Monitoring (PM) data collected from six international optical networks, we find that it is possible to forecast LOS events with good precision 1-7 days before they occur, albeit at relatively low recall, with supervised machine learning (ML). Our study covers twelve facility types, including 100G lines and ETH10G clients. We show that the precision for a given network improves when training on multiple networks simultaneously relative to training on an individual network. Furthermore, we show that it is possible to forecast LOS from all facility types and all networks with a single model, whereas fine-tuning for a particular facility or network only brings modest improvements. Hence our ML models remain effective for optical networks previously unknown to the model, which makes them usable for commercial applications.
\end{abstract}


\section{Introduction}
Optical networks form the backbone of the global information and communication infrastructure, which supports a huge ecosystem of technologies and services. Thus, the high availability of the network infrastructure is critical both economically and socially. Although today's networks are remarkably reliable, Loss of Signal (LOS) still occurs. When it happens, either (1) the network protects itself automatically or (2) the LOS propagates to cause signal interruption. This latter situation degrades service quality for end-users, incurs labor and maintenance costs for the network operator and can even have catastrophic effects. For instance, in 2016, Southwest Airline canceled more than 2,000 flights and lost tens of millions of dollars due to a massive network system failure caused by a faulty router~\cite{boamah2019southwest, dallas2016southwest}. Predicting LOS events before they occur enables proactive actions to prevent such network outages.

Autonomous networking is the vision pursued by world-leading network technology companies, including the Adaptive Network~\cite{ciena_adaptive} proposed by Ciena, the Autonomous Driving Network proposed~\cite{huawei_adn} by Huawei, the Digital Network Architecture proposed~\cite{cisco_dna} by Cisco, the self-driving network proposed by Juniper~\cite{Juniper_selfdriving}, and the Zero-Touch Network~\cite{ericsson_zerotouch} proposed by Ericsson. This activity is largely focused on automating reactive processes, but a few applications such as Ciena's Network Health Predictor (NHP)~\cite{ciena_nhp} and Juniper's HealthBot~\cite{juniper_healthbot} can also forecast network events before they occur to strengthen service reliability.

\begin{figure}[!htb]
	\centering
	\subfigure[Time series trend of PM QAVG]{
		\begin{minipage}[t]{0.95\textwidth}
			\centering
			\includegraphics[width=16cm]{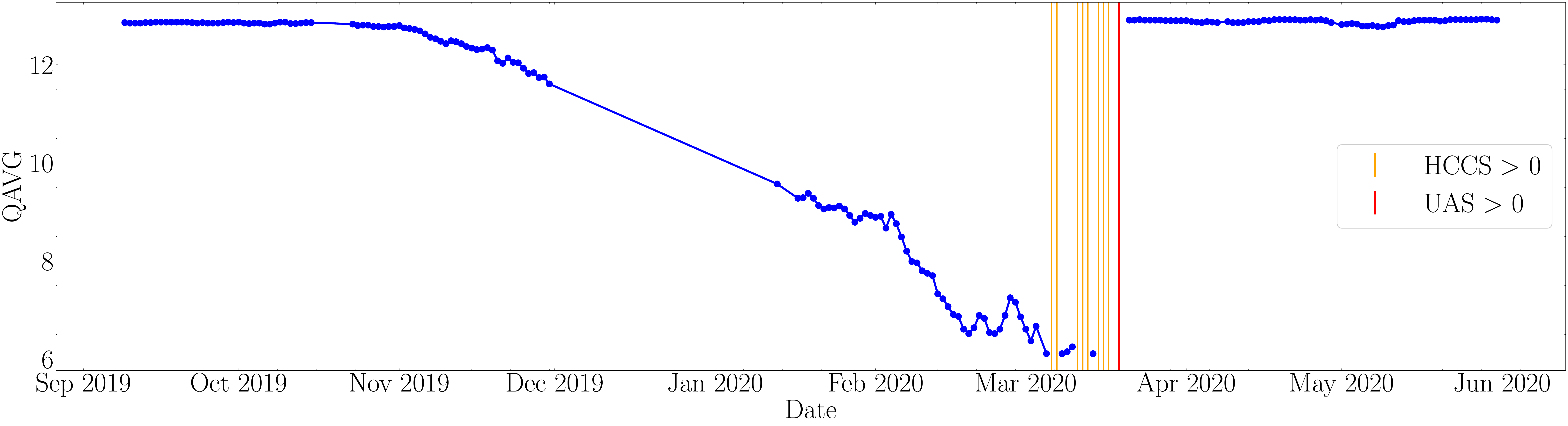}
			\label{fig:QAVG}
		\end{minipage}
	}
	\subfigure[Time series trend of PM QSTDEV]{
		\begin{minipage}[t]{0.95\textwidth}
			\centering
			\includegraphics[width=16cm]{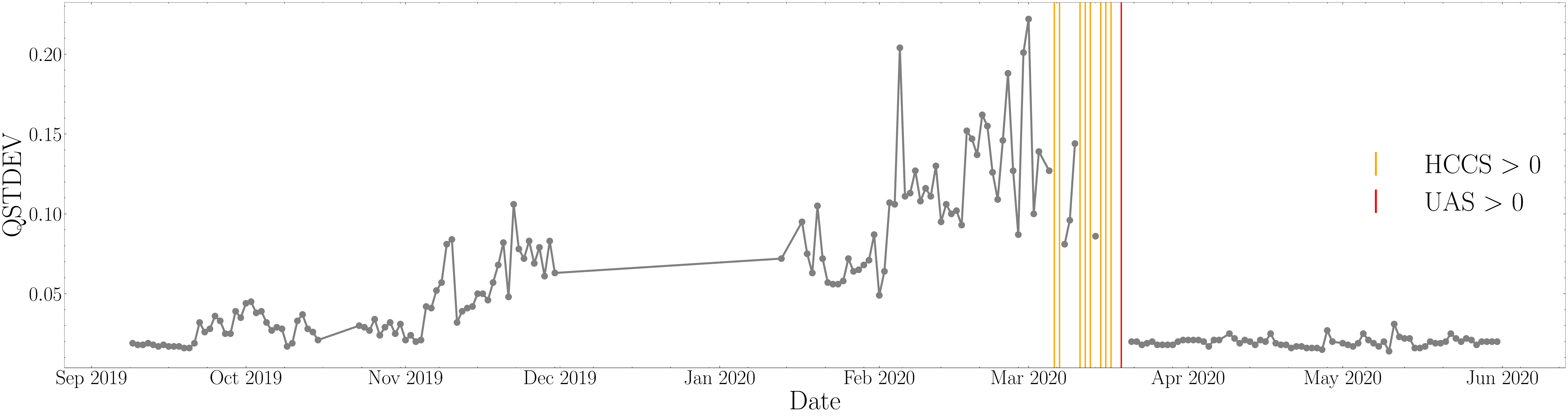}
			\label{fig:QSTDEV}
		\end{minipage}
	}
	\caption{
		A typical example of predictable Imminent Loss of Signal (ILOS) from a production network. Figures~\ref{fig:QAVG} and~\ref{fig:QSTDEV} show signal quality (QAVG) and stability (QSTDEV), respectively, over a time period ranging from September 2019 to May 2020. Here $Q$ is a factor representing the signal quality. QAVG is the average value of Q on that day and QSTDEV is the standard deviation of Q value on that day. Orange vertical lines indicate days with transient LOS occurring during a few seconds (HCCS $\in [1,60]$ sec). The red vertical line indicates an outage where the LOS lasted over an hour (UAS > 3600 sec). This LOS event could have been predicted beforehand given these PMs.
	}
	\label{fig:ILOS_PM_distribution}
\end{figure}

In optical networks, service-impacting LOS often comes from intrinsically unpredictable fiber cuts. But other root-causes like equipment aging, loose connectors, or system mis-configuration can create predictable LOS events. Hence it is possible to forecast some LOS in advance, though not all of them. Figure~\ref{fig:ILOS_PM_distribution} shows one compelling example of predictable LOS reported from the 100G line receiver port of a Ciena 6500 device in a production network. In the left-most part of the graph, the signal quality is good (high QAVG) and stable (low QSTDEV). Here Q is a factor representing the signal quality. QAVG is the average score of Q (reflecting the signal quality) and QSTDEV is the standard deviation of Q (reflecting the signal stability). Then, over a period of several months, the signal quality becomes incrementally unstable and degraded. In March 2020, orange lines indicate that the port experiences transient LOS for a few seconds. Finally, a red vertical line indicates an outage where the LOS lasts over an hour before the issue gets fixed and the ports re-start reporting good and stable signal quality again. From these findings, we speculate that machine learning (ML) models may be able to forecast Imminent Loss of Signal (ILOS) from receiver ports in multi-vendor packet-optical networks that report similar data.

Recently, ML has been applied in solving telecommunication problems. The exploratory experiments by C\^ot\'e in~\cite{cote2018machinelearning} validated the speculation that the random forest model is applicable to abnormal element detection in optical networks. However, it used laboratory data and it lacked the ability to predicate ILOS for 1 to 7 days in the future. In this work, we focus on six sets of real-world data collected from large commercial optical networks geographically distributed around the world. 

Other researchers have reported success in using ML to reduce nonlinear phase noise~\cite{Pan2015nonlinear, Zhuge2019nonlinear}, estimate quality of transmission~\cite{Tanimura2019monitoring,Seve2021,Rottondi2018}, detect anomalies~\cite{Shahkarami2018failure_detection, Chen2019anomaly_detection, Rafique2018Cognitive}, forecast equipment failures~\cite{yan2019alarm, Zhuang2020AlarmPrediction,wang2017failure, Christodoulopoulos2019ORCHESTRA,Morais2021}, and enable AI-based routing~\cite{Zhong2019Routing} or self-optimizing networks~\cite{zhao2018soon,Zami2020}. Building on this valuable work, we aim to develop a general engineering framework that can handle telecommunication data for production-grade applications. This real-world data comes with practical challenges for ML best practices today, notably: (1) it is unlabeled, (2) it has imperfect data-quality, and (3) it is not readily available in large sets.

To overcome these challenges, we have developed a framework that can:

\begin{enumerate}
	\item \textbf{Encode domain knowledge} \\
	Automatic rules developed by experts can identify problematic network elements (NE) from performance monitoring (PM) data after the fact. We use these rules to process historical datasets and label the status of each NE over time. Then we use labeled data samples to train supervised ML models that predict good $\rightarrow$ bad transitions before they occur.
	
	\item \textbf{Handle imperfect data quality} \\
	Our input data is not always continuous and homogeneous. While not ideal, this is beyond our control and we have to be pragmatic about it. Notably, we can miss data because of zero-suppression, because of error conditions, or because the data was never collected. As a result, our overall matrix of features has a sparse structure with a high missing rate (>70\%). Traditional imputation methods do not perform well in this situation, which led us to develop sophisticated methods to learn from the data present without being biased by the data missing.
	
	\item \textbf{Combine datasets from diverse sources} \\
	Deep learning methods generally benefit from the largest possible data samples, especially for solving needle-in-the-haystack problems like predicting optical network outages. Nevertheless, Network Management Systems in production typically keep data for only a few days or weeks. Hence we have collected data multiple times from multiple networks to accumulate big datasets. However, commercial optical networks are complex and diverse, hence we put in place a substantial machinery to combine data from all sources into a single \emph{mega-dataset}.
\end{enumerate}

Equipped with the above, we have been able to train supervised ML models that can predict ILOS 1-7 days before they occur in commercial optical networks. Furthermore, we show that this task can conveniently be accomplished with a single ML model that covers all Layer 1 and 2 devices across a full packet-optical network.

In Section~\ref{related-work}, we review the prior work related to this topic. In Section~\ref{methodology}, we describe our data pre-processing and ML methodology. In Section~\ref{sec-experiments}, we describe the various ML experiments that we performed during our analysis, and in the same section, we present our final results. Finally, in Section~\ref{sec-conclusion}, we summarize our main conclusions and give an outlook on the next steps.

\section{Related Work} \label{related-work}
\subsection{Handling Missing Values}
Missing values are ubiquitous in time series datasets because of communication interruption or sensor malfunction during data collection, especially for industry data. Imputing missing values with mean, median or fixed values such as 0 are basic methods to process missing data. Most of such simple imputation methods impose strong assumptions on missing data. For example, zero imputation presumes missing values are zero suppression. However, a high missing rate makes it very challenging for these conventional imputation methods to be effective because a high missing rate results in diverse missing patterns, making these assumptions hard to meet~\cite{cao2018brits}.

In recent years, some researchers attempted to solve this problem of missing values with ML. XGBoost proposed by Chen et al.~\cite{chen2016xgboost} can directly process input with missing values. As a tree-based algorithm, XGBoost has a learning mechanism called sparsity-aware split finding to assign a default direction to each tree. When XGBoost encounters missing values, the branch takes the default direction to continue the decision path. Marek Smieja et al.~\cite{smieja2018missing} model the uncertainty on missing attributes by probability density functions using the Gaussian mixture model (GMM), then replace the typical neuron's response in the first hidden layer by its expected value. Recurrent neural networks (RNN) are always popular in the time series topic, and this missing value problem is not an exception. Che et al.~\cite{che2018grud} propose GRU-D, which introduces a decay mechanism into a GRU so that the influence of input variables is decayed over time if the variable has been missing for a while. Luo et al.~\cite{luo2018gan} propose the Gated Recurrent Unit for data Imputation (GRUI) to model the incomplete time series and integrate the GRUI into a generative adversarial network (GAN) to treat imputation as a data generating task. BRITS (Bidirectional Recurrent Imputation for Time Series) proposed by Cao et al.~\cite{cao2018brits} directly learns missing values in a bidirectional recurrent dynamical model, where imputed values are treated as variables of RNN. At the time of writing, BRITS achieves the state-of-the-art (SOTA) performance on the PhysioNet Challenge 2012 dataset~\cite{silva2012physio}, a time-series health-care dataset with 78\% of the values missing~\cite{cao2018brits}. Hence BRITS is the neural network model we apply to solve missing data problem in this work.

\subsection{Detecting and Forecasting Anomalies in Optical Networks}
Recently, ML has been utilized for anomaly detection and forecasting in optical networks. To detect anomaly, Chen et al.~\cite{Chen2019anomaly_detection} employ an unsupervised density-based clustering algorithm to analyze monitoring data patterns and then utilize a self-taught mechanism to transfer learned patterns to a supervised data regression and classification module. Rafique et al.~\cite{Rafique2018Cognitive} build a proactive fault detection (PFD) engine for autonomous anomaly detection. The PFD engine firstly use the generalized extreme studentized deviate (ESD) test to identify all potential faults and then send them into a neural network to classify true anomalies. Shahkarami et al.~\cite{Shahkarami2018failure_detection} define a detection framework for Bit Error Rate anomaly and propose a ML method to discriminate different sources of soft failure. 
To predict the risk of an equipment failure, Zhilong Wang et al.~\cite{wang2017failure} propose a performance monitoring and failure prediction method based on support vector machine (SVM) and double exponential smoothing (DES). They use DES to converge values of features indicating the status of equipment and then predict future values sent into the SVM model to classify if the facility will fail in the near future. With regards to labeling, they apply a threshold on the indicator 'Unusable Time' and assume equipment has failed if this indicator's value exceeds the threshold. The ORCHESTRA network presented in~\cite{Christodoulopoulos2019ORCHESTRA} is developed to predict, detect, and diagnose health issues automatically, for example, soft failures like quality of transmission (QoT) degradation. Self-Optimizing Optical Networks (SOON) proposed by Zhao et al.~\cite{zhao2018soon} is a network architecture based on software-defined networking with ML for applications with tidal traffic forecasting, alarm prediction, and anomaly action detection. Due to inferior data quality, Zhuang et al.~\cite{Zhuang2020AlarmPrediction} leverage a GAN to augment their dataset and train a neural network on the augmented data to predict alarm in optical networks.

The above methods do not specifically focus on handling missing data, the small size of data samples, or combining imputation with ML. For example, in Zhilong Wang's experiments~\cite{wang2017failure}, their dataset has only 15 features and 14,080 samples without missing data, and sample classes also get balanced intentionally. There is even no need to distinguish facility types in their data. In the research of Yan et al.~\cite{yan2019alarm}, they demonstrate a dirty-data-based alarm prediction method for their specific network structure based on SOON. Data is collected from commercial large-scale optical networks and has characteristics similar to the data used here, such as imbalanced classes and missing data. However, their work is based on the SOON architecture, not generally applicable to other networks. Additionally, they work on short-term prediction, which takes in data collected from the past 3 hours to predict alarms in the future 15 minutes. Such short-term prediction does not leave enough time for operators to respond and perform maintenance.

\subsection{Applying Transfer Learning to Optical Networking}
Transfer learning transfers the learned knowledge from the source domain to the target domain. It emerges to tackle the problem of building deep learning models in the domain of interest by leveraging the sufficient data in another related domain~\cite{pan2010transfer}. In the work of Yu et al.~\cite{yu2019transfer}, transfer learning is adopted to enable applying their neural network models to different optical systems with only a small size of new training samples. Xia et al.~\cite{xia2019transfer} propose a deep neural network to estimate optical-signal-to-noise ratio (OSNR). They leverage transfer learning to remodeling if system parameters change. Their method is able to reduce the training time by a factor of four and requires only one-fifth the training set size without any performance loss. To detect new cyber-attack behaviors in a network, Zhao et al.~\cite{Zhao2019TransferLF} propose a clustering-enhanced transfer learning (CeHTL) approach to find out relations between new attacks and known attacks. Azzimonti et al.~\cite{azzimonti2020transferleraning} utilize Gaussian process to estimate bit error rate in an unestablished lightpath. In their work, transfer learning is adopted to further optimize the accuracy on small-sized datasets.

\section{Methodology} \label{methodology}
Our methodology consists of three aspects: (1) data pre-processing and labeling, (2) core modeling with missing data handling embedded into the learning model, and (3) transfer learning across networks.  
For data pre-processing, we produce port-level time-series datasets to represent the status of one port on one day. The labels of data samples are generated by our rule-based method. For modeling, we formalize the ILOS prediction problem into a binary classification modeling task solved with supervised ML. Furthermore, we compare classifier algorithms with three different approaches to missing data: random forest~\cite{leo2001rf}, XGBoost~\cite{chen2016xgboost} and BRITS~\cite{cao2018brits}.

For transfer learning, we seek to improve the learning performance on small datasets that lack sufficient training data by leveraging a well trained model with a similar feature space. Our solution is to combine all datasets from six networks into one mega-dataset. The mega-dataset enables these networks to share feature space across all datasets. Models pre-trained on the mega-dataset learn the general feature representation of ILOS signatures from all datasets, akin to knowledge sharing. Our experiments observe that such general ILOS representation improves the performance up to 36.4\% as compared to the deep learning model on networks with small datasets and up to 75.8\% as compared to XGBoost.

\subsection{Data Pre-processing}
In our data pre-processing methodology, we first introduce properties of our collected data and then illustrate our six pre-processing steps in details.

\subsubsection{Dataset Description} \label{dataset-desp}
The datasets used for this research consist of daily-binned Performance Monitoring metrics (PMs) collected from equipment in production network environments, spanning layers 1 and 2. Each device typically has multiple ports, where each port reports PMs for some facility, where, in turn, a facility is a low-level object mapping to a physical or virtual component or service of the Optical Transport Network (OTN) such as e.g. Ethernet (ETH) or Optical Transport Module (OTM).

Our analysis uses every PM reported by the port as a feature input to the model. These PMs report a wide variety of information related to the signal quality, describing e.g. the number of input and output frames, number of code violations, optical power, delay measurements, number of errored seconds, failure counts, number of forward error corrections, signal quality ("Q") and other metrics related to low-level errors in the signal.

An overview of our datasets is shown in Table~\ref{tbl1}, where datasets for Network1 to Network6 are sorted by values of positive sample rate in incremental order from left to right. The mega-dataset is the one combining all 6 datasets for transfer learning. It is worth noting that classes in our datasets are imbalanced. The overall positive sample rate in the mega-dataset is around 10\%. In Network1, the positive sample rate is less than 5\%. A small number of positive samples is consistent with the observation mentioned above that equipment failures are rare in optical networks.

As summarized in Table~\ref{tbl1}, we consider twelve facility types and nine protocols to characterize the state and performance of the entire network, from physical equipment (layer-0) to optical (layer-1) and Ethernet (layer-2) signals. 

\begin{table}[ht]
	\centering
	\caption{Details of the six network datasets used in this paper, sorted by positive sample rate in ascending order from left to right. Each dataset includes all 12 facility types of layer-1 and layer-2 ports. The right-most column summarized the "mega-dataset", which is a combination of all six network datasets. The missing rate of the mega-dataset is the largest due to the fact that some networks have features that do not exist in others, resulting in nearly empty columns when merging the network datasets into the mega-dataset.}
	\resizebox{160mm}{!}{
	\begin{tabular}{c|cccccc|c}
		\toprule
		& Network1 & Network2 & Network3 & Network4 & Network5 & Network6 & Mega-dataset\\
		\midrule
		Number of protocol types & 6 & 9 & 8 & 6 & 8 & 8 & 9\\
		\midrule
		Time range (days) & 139 & 222 & 204 & 299 & 301 & 208 & 1,373 \\
		\midrule
		Number of ports &  3,000 & 19,000 & 5,000 & 35,000 & 16,000 & 16,000 & 94,000 \\
		\midrule
		Number of samples & 394,158 & 3,540,472 & 762,268  & 6,140,585 & 3,140,132 & 2,478,716 & 16,456,331 \\
		\midrule
		Number of features & 76 & 83 & 113 & 72 & 115 & 91 & 125 \\
		\midrule
		Missing rate & 75.2\%   & 72.4\%   & 79.2\%   & 70.1\%   & 77.6\%   & 79.8\%    & 82.2\%  \\
		\midrule
		Positive sample rate & 4.3\%  & 5.7\%    & 5.9\%    & 8.6\%    & 15.8\%   & 17.0\%    & 10.4\%   \\
		\bottomrule
	\end{tabular}
	}
	\label{tbl1}
\end{table}

\subsubsection{Pre-processing Steps}
\begin{figure}[ht]
	\centering
	\includegraphics[width=16.1cm]{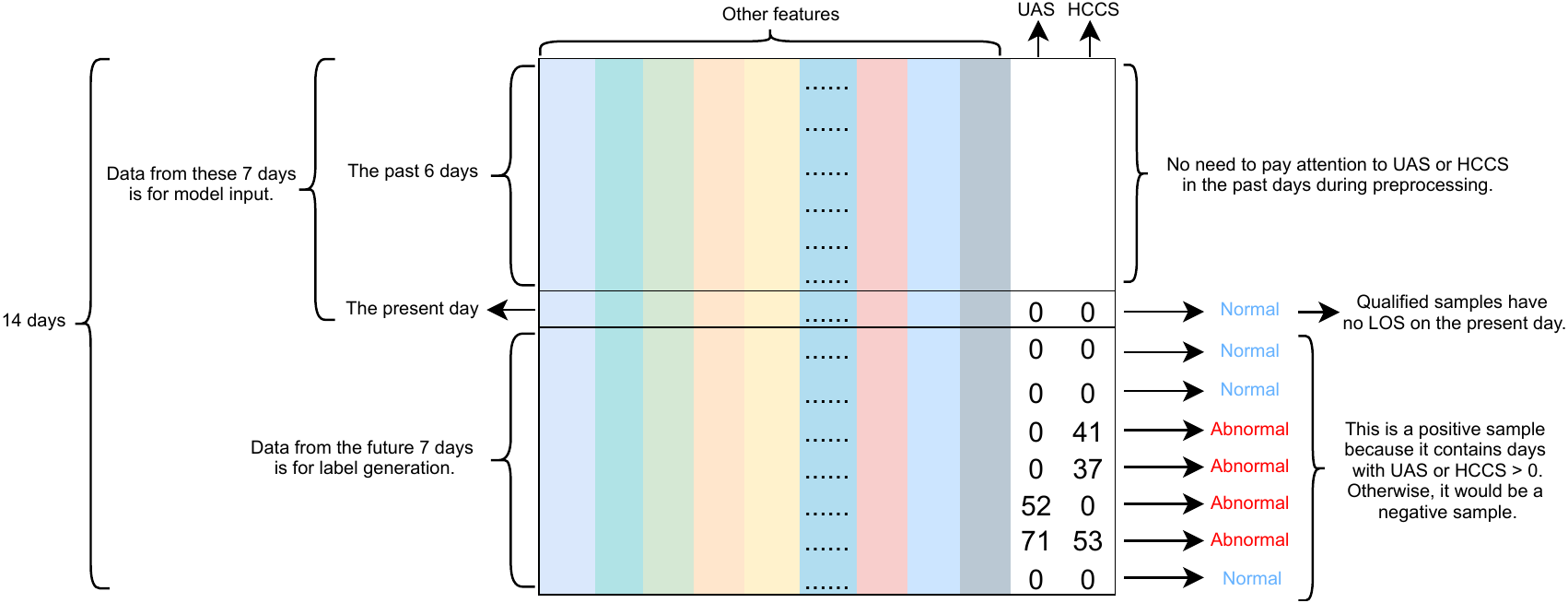}
	\caption{Overview of a qualified complete sample and labeling method.}
	\label{fig:Labeling_Overview}
\end{figure}

There are six steps to pre-process our data, listed as follows:

\textbf{1. Re-organize samples to port-level.}
Considering that our experiments need port-level datasets, raw data needs to be re-organized into port-level. Facility-level samples are grouped by their port IDs and the collection timestamps. This means these facility samples are collected from the same port on the same day. These facility-level samples are merged into one line to compose port-level samples. In the merging operation, only maximum values are kept if one feature has multiple values because of more than one facility in the port. Facility types are converted into one-hot encoding. As a result, each sample in the datasets can represent the status of one port on one day.

\textbf{2. Produce time-series data for deep learning.}
To generate time-series datasets for deep learning, samples of each port are sorted by collection timestamps firstly. Considering that the collection process may be interrupted for different reasons, missing days are first filled, with all PM values as NaN (Not a Number), so that all samples of one port are time-continuous without interruption. As a result, a 14-day window slides on the chronological data of each port to produce time-series samples. Each produced sample contains data collected from one port in 14 consecutive days. The generated samples contain some days that are filled artificially as NaN.

Each sample is of a 14-day span that is further divided into three parts, namely: (1) \textit{ the present day} as the seventh record; (2) \textit{the past six days} as the first six records; (3) \textit{future seven days} as the last seven records. The first seven days of all the samples are used for model training, while the future days of all the samples are used for labeling. The labels indicate whether the port encounters an ILOS on each day of the last seven days. The model learns the features in the first seven days about the status of the port and then predicts if an ILOS may occur on the port in the next 7 days. 

\textbf{3. Generate labels.}
We rely on data labels to interpret PM values into binary category positive (with ILOS) or negative (without ILOS), which is necessary for status analysis of facilities or ports. 
However, accurate labels are difficult to obtain in large telecommunication network datasets because of the specialized domain expertise that is required~\cite{cote2018machinelearning}. 
In this work, we have used rules provided by optical network experts to label historical data automatically. The labeling logic uses the fact that a port is generally suffering from LOS when it experiences errors for a sufficient period of time. For our 6500 devices, this error metric is well-represented by two PMs: Unavailable Seconds (UAS) and High Correction Count Seconds (HCCS). UAS counts the number of seconds during which a facility was unavailable and unable to perform its task, while HCCS counts the number of seconds per day where the number of errors to be corrected by "forward error correction" techniques exceeds some threshold. Indeed, for a smaller dataset for which alarm data was available, we have verified empirically that these two PMs are highly correlated with LOS alarms emitted by the devices. Thus, if a sample contains UAS or HCCS larger than zero on any of its future 7 days, we label the sample as positive, indicating there is a LOS in the future. We plot Figure~\ref{fig:Labeling_Overview} to further show a concrete example of how our labeling method works.

\textbf{4. Filter out defective samples.}
Samples that meet the following conditions are filtered: (1) not carrying traffic in the present day. There are protocol features in PMs indicating whether the port is carrying traffic. If it is not, we assert this port is not working and filter such a sample; (2) having LOS in the present day may already trigger maintenance operations. Therefore predicting ILOS for such samples is a special case. To focus on the prediction of ILOS in general, we filter out the data samples with LOS in the present day; (3) no data for all of the first 7 days or future 7 days. In this extreme case with no data for all of the first and last 7 days, rather than filling in data, we filter out such samples. 

\textbf{5. Split into different parts.}
We split our datasets into 3 parts, including the training set, the validation set, and the test set. Samples in datasets are sorted according to date timestamps. Then the partition of the datasets is approximately training set 70\%, validation set 10\%, and test set 20\%.

\textbf{6. Normalize data.}
For neural network models depending on gradient descent, feature normalization is necessary. We apply Z-score normalization to feature vectors.

\subsection{Model Details and Implementation} \label{model-implementation}
The core principle of our solution is to augment ML models with missing data handling. We examine and expand each representing model as follows. Random forest is a classical boosting tree algorithm that does not have a default mechanism for handling missing data. Hence we apply two conventional imputation methods of zero imputation and median value imputation to process missing values. XGBoost, on the other hand, can handle missing data by default. BRITS is an RNN-based deep learning algorithm. Missing values in feature vectors are located by missing masks and then get imputed by zeros. Concatenations of imputed feature vectors and missing masks are input into BRITS to let BRITS learn missing values by itself.

\subsubsection{Random Forest as the Baseline Model}
Random forest is adopted in the work of C\^ot\'e~\cite{cote2018machinelearning} to discriminate states of network elements as being 'Normal' or 'Abnormal'. For simplicity, random forest is also adopted in our methodology to observe the learning effects of our proposed models. In addition, we expand the random forest model with two conventional imputation methods to process missing values in the input data. One method uses fixed value (0) imputation and the other method adopts the median value imputation. The comparison between these two variants to the random forest model shows how selecting different traditional imputation methods can affect model performance on our datasets. Considering that random forest is not a time-series model, each sample containing 7-day sequence data (namely 7 rows) is expanded into 1 row before being input into the model. 

Random forest in our experiments is the implementation in the python library scikit-learn~\cite{sklearn}. For this tree-based model, we find that the parameter \textit{number of trees} has the greatest impact on our results. Therefore, we search for the most appropriate parameter value in a range from 100 to 3,000 in intervals of 100 and validate each one's performance on the validation set to pick out the best.

\subsubsection{XGBoost}
XGBoost is a scalable end-to-end tree boosting system. XGBoost has a wide range of applications in ML challenges and has achieved strong results effectively. We consider XGBoost because it is a sparsity-aware algorithm~\cite{chen2016xgboost}. As discussed in Section~\ref{related-work}, XGBoost assigns missing values to the default direction to continue the decision path. Moreover, the optimal default direction is found by trying both directions in a splitting node so that XGBoost can minimize the training loss. Therefore, missing values in the input of XGBoost can just be left as NaN for XGBoost to handle by itself.  The input of XGBoost also needs to be reshaped into 1 row.

In our work, the implementation of XGBoost is from the open-source package mentioned in the original paper~\cite{chen2016xgboost}. For XGBoost, we also perform parameter searching on \textit{number of trees} to optimize the performance.

\subsubsection{BRITS: Bidirectional Recurrent Imputation for Time Series}
BRITS model is proposed by Cao et al.~\cite{cao2018brits} to handle missing values in multivariate time series data. BRITS learns missing values in its bidirectional neural network system without imposing any specific assumptions on the data generating process, which means it is generally applicable to time series imputation tasks. As a multi-task learning model, BRITS is trained on both the imputation task and the classification task. BRITS is able to make imputations for missing values and solve classification at the same time. 

\textbf{BRITS model structure.} BRITS model is a bidirectional model. It consists of 2 RITS (Recurrent Imputation for Time Series) models. One RITS takes in data in the forward direction and the other RITS takes in data reversed, namely in the backward direction. In the training period, both models' outputs should agree with each other after the backward one’s output gets reversed, or the model gets a penalty that is the consistency loss. The consistency loss is the discrepancy between imputations from 2 models, which is measured by mean absolute error (MAE). 

\begin{figure}[ht]
	\centering
	\includegraphics[height=277pt]{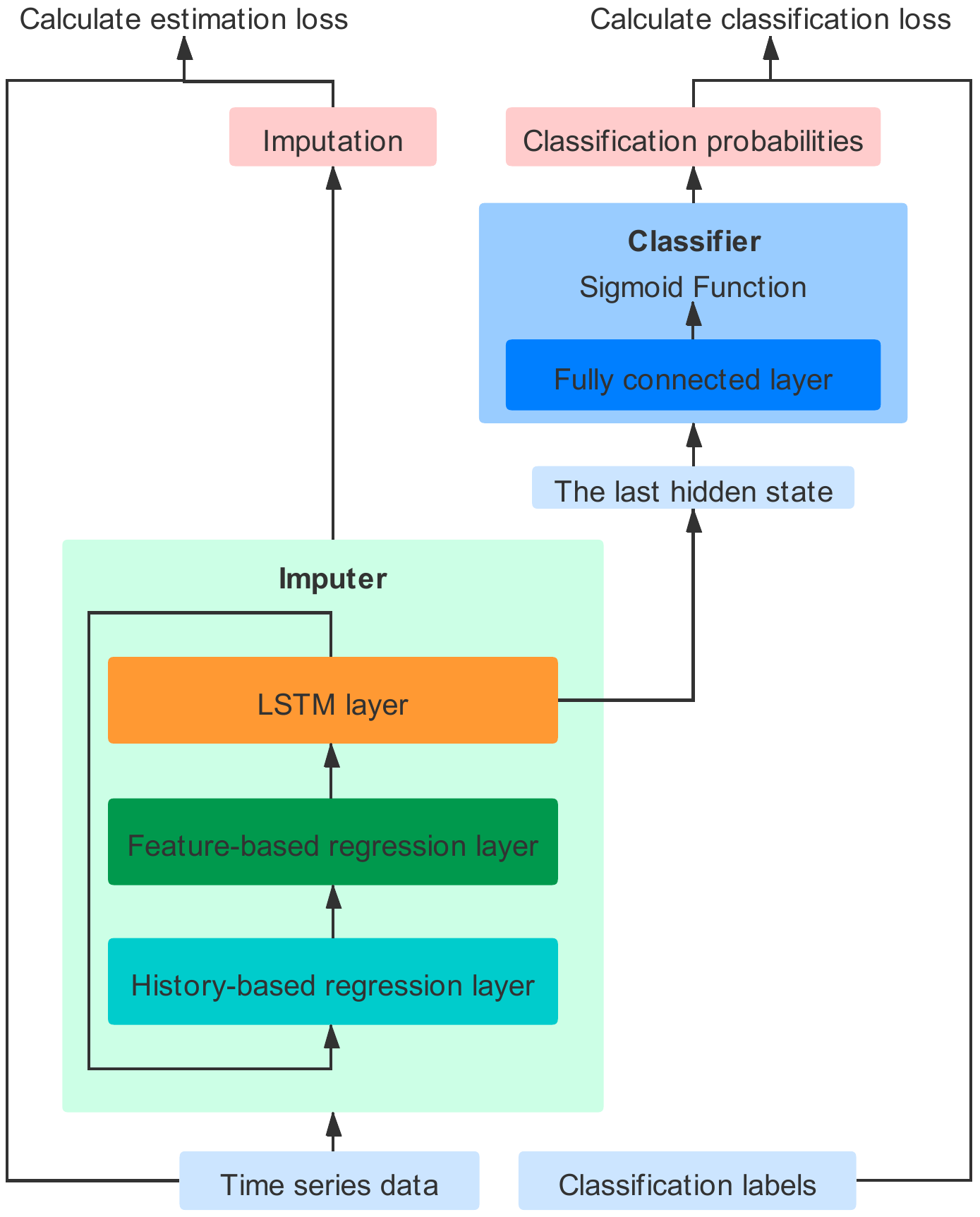}
	\caption{Structure overview of the Recurrent Imputation for Time Series (RITS) neural network model.}
	\label{fig:RITS_overview}
\end{figure}

As shown in Figure~\ref{fig:RITS_overview}, each RITS model is made of 2 parts. One part of a RITS is the imputer, which consists of a LSTM (Long Short-Term Memory)~\cite{hochreiter1997lstm} layer and regression layers to learn the representation of features across time and correlation between features. An imputation task matches observed feature values with imputations generated by the imputer and calculates MAE between them to generate the estimation loss. The other part of a RITS is the classifier that has a fully connected layer. This fully connected layer takes the last hidden state output from the LSTM layer in the imputer to produce \textit{logits}. Then the \textit{sigmoid} function converts the \textit{logits} into classification probabilities. Errors made by the classifier result in the classification loss. Considering that BRITS consists of 2 RITS models and each model has its results for losses, imputations, and classification probabilities, the results that BRITS finally yields are averaged values of each RITS model. The total loss in BRITS is the sum of the above mentioned consistency loss, estimation loss and classification loss. 

\textbf{BRITS data input.} Data input into BRITS consists of 4 parts: feature vectors, missing masks, time gap vectors, and classification labels. It is worth mentioning that in the original work of Cao et al.~\cite{cao2018brits}, they randomly masked out an extra 10\% data in their datasets that was held out for comparing BRITS' imputation ability with other models in the test period. In this paper, BRITS is utilized to help us handle missing data, hence such extra-masked data do not exist in our input. Feature vectors contain all processed feature values. Missing masks indicate missing values in feature vectors. As shown in Equation~\ref{equation2}, if $x^d_t$ is missing that means feature $d$ is missing at time $t$, then $m^d_t$, the mask value of feature $d$ at time $t$, is set as 0. If $x^d_t$ is observed, $m^d_t$ is set as 1. Time gap vectors contain each feature's time gap from the last observation to the current timestamp. They are calculated from missing masks and are used to generate temporal decay factors to decay hidden states in the LSTM layer. Equation~\ref{equation3} illustrates the calculation of time gap vectors $\delta_t$ from the last observation to the current timestamp $s_t$ in detail. The working principle of temporal decay factors is that if values are missing for a long time, then the last observations have less correlation with values at the current time step. As a consequence, corresponding hidden states are expected to be decayed more.

\begin{equation}
	m^{d}_{t} =
	\begin{cases}
		0 & x^d_t \: is \: missing \\
		1 & x^d_t \: is \: observed \\
	\end{cases}
	\label{equation2}
\end{equation}

\begin{equation}
	\label{equation3}
	\delta_t = 
	\begin{cases}
		$$s_t$$ - $$s_{t-1}$$ + \delta^{d}_{t-1} & t>1, m^{d}_{t-1}=0 \\
		$$s_t$$ - $$s_{t-1}$$					 & t>1, m^{d}_{t-1}=1 \\
		0										 & t=0
	\end{cases}
\end{equation}

\begin{figure}[!htb]
	\centering
	\includegraphics[width=16cm]{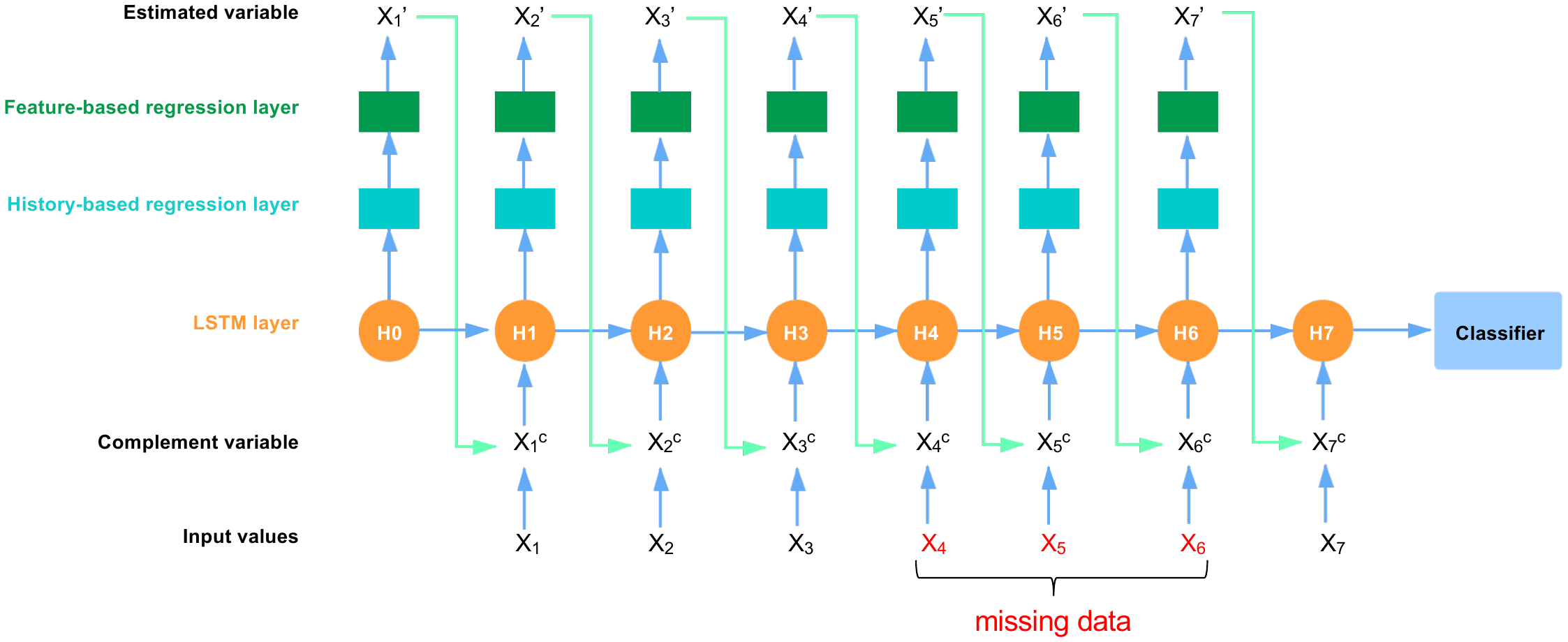}
	\caption{Detailed processing steps of the RITS NN architecture from the forward direction (from $X_1$ to $X_7$). For the backward RITS, data flows from $X_7$ to $X_1$.}
	\label{fig:BRITS_steps}
\end{figure}

\textbf{BRITS processing flow.} We plot Figure~\ref{fig:BRITS_steps} to display detailed processing steps in the forward RITS that takes in data in the order from $X_1$ to $X_7$. Likewise, the backward RITS processes data in the order from $X_7$ to $X_1$. For each step of data, its hidden state is sent into the history-based regression layer to produce the estimation of missing values based on history values. The first step has no hidden states before it, hence $H_0$ here is initialized as a 0 vector. The output of the history-based layer is sent into the feature-based regression layer to produce the feature-based estimation. When calculating the estimation of missing values in one feature, the feature-based regression layer takes the correlation between this feature and others into consideration. Subsequently, the model combines the history-based estimation and the feature-based estimation by learned weights to form estimated variable $X'$. $X'$ is used to impute missing values in $X$ to form the complement variable $X^c$. The LSTM layer takes $X^c$ to process temporal information and produce a hidden state for the next step. Such a cycle repeats for the whole time series. Finally, the classifier takes the last hidden state to produce the probability of ILOS. Our BRITS implementation is available in the GitHub code repository~\footnote{https://github.com/WenjieDu/PyPOTS}. We make modifications to fit our datasets. In our experiments, BRITS obtains the best performance when the RNN hidden size equals 256. Therefore, we fix this most important hyper-parameter as 256 for all BRITS models used in this work. 

\subsection{Transfer Learning}
Transfer learning is adopted in our work to help our models obtain better performance on small datasets. We design the transfer learning with two stages: pre-training and fine-tuning. In practice, the whole transfer learning methodology is usually applied on neural networks only, such as BRITS. Although random forest and XGBoost have no way to get tuned like BRITS, they still can be pre-trained to learn general knowledge across our network datasets. Therefore, pre-training here can also be considered as knowledge sharing, which shares the general representation of ILOS signatures across networks.

To collect the general representation of data,  we merge six datasets into one, called the mega-dataset. Training sets, validation sets, and test sets from each network dataset are  merged into the corresponding mega-training set, the mega-validation set, and the mega-test set. For features that are different between networks, we take the union of these features.  Consequently, the feature number of the mega-dataset is 125, as shown in Table~\ref{tbl1}. We also keep a column to record which network the sample belongs to. This column is not a feature and not visible to models, which is only used to filter out other networks' data while fine-tuning models on the network-specific dataset.

\begin{figure}[!htb]
	\centering
	\includegraphics[width=16cm]{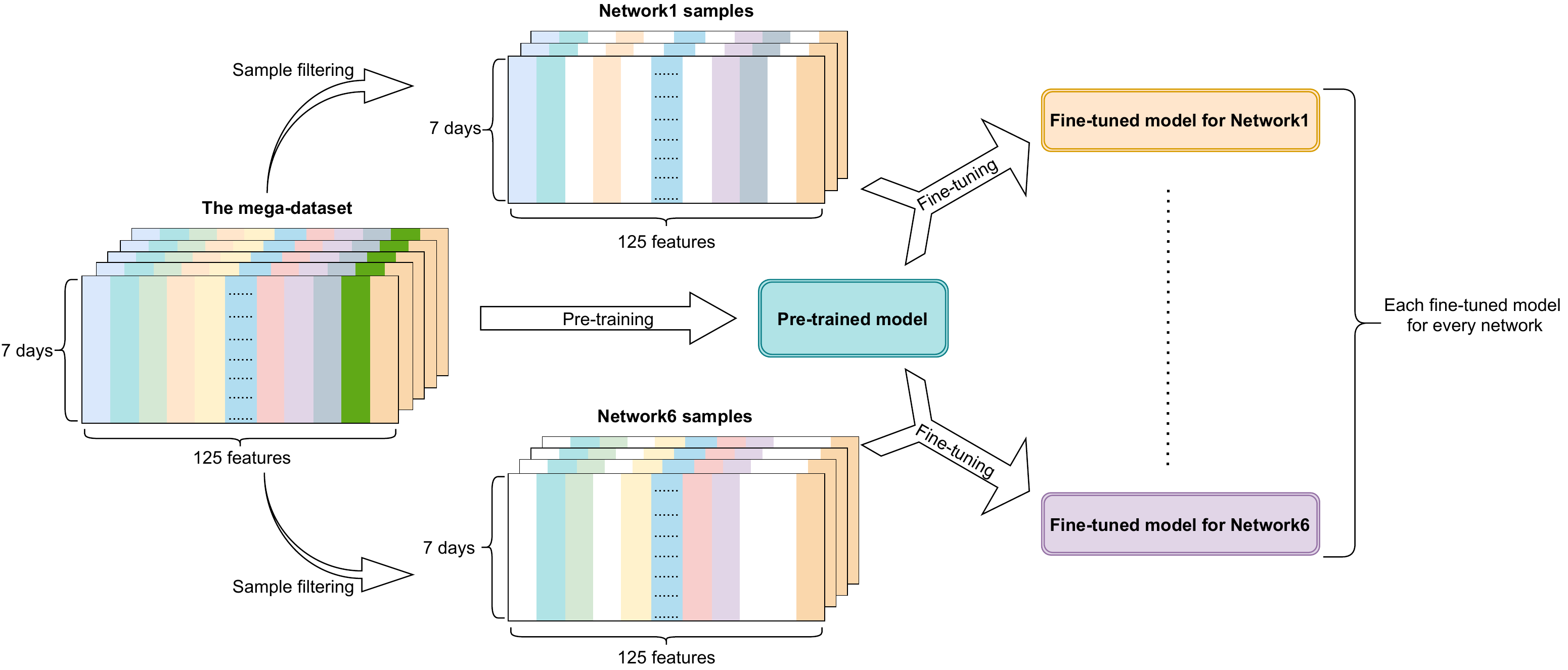}
	\caption{Overview of our transfer learning methodology. Combining all six network datasets, our "mega-dataset" has 125 features for each of the past 7 days. Taking the datasets for Network1 and Network6 as an example, some features in Network1 may not exist in Network6, and vice versa (white/blank columns in the figure). Our model is initially pre-trained on the mega-training set and subsequently fine-tuned on network-specific samples to generate fine-tuned models for each network, resulting in a total of 6 models.}
	\label{fig:transfer_learning}
\end{figure}

As shown in Figure~\ref{fig:transfer_learning}, BRITS is firstly pre-trained on the mega-dataset to learn general representation and then fine-tuned on single-network datasets to learn network-specific knowledge. In pre-training, BRITS is just trained on the mega-dataset rather than single-network datasets, and the learning rate is set as $10^{-3}$. 

Fine-tuning is only to keep a specific network's data and let pre-trained BRITS adjust its parameters on these network-specific samples. The learning rate in the fine-tuning stage is halved to $5 \times 10^{-4}$. The general practice of fine-tuning is to tune the classifier but keep other parts frozen. In our case, this means freezing parameters in the imputer and only retraining the classifier of BRITS on network-specific datasets. In another strategy, we also expand the fine-tuning to the imputer, which means parameters of the entire BRITS model are adjusted on network-specific data during the fine-tuning stage. For random forest and XGBoost, these two models only have the pre-training stage without the fine-tuning stage.

\section{Experiments} \label{sec-experiments}
In this section, we design three experiments to validate our methodology. First we compare across models trained on individual datasets. Second, we discuss the feasibility of transferring the general representation of ILOS signatures across networks. In particular, we observe how much improvement transfer learning can bring to our models on small datasets. Third, we benchmark the performance of our models on the two main commercial use-cases of NHP to show our work is valuable in practical applications.

It is worthwhile to mention that, in this section, models mentioned in italic font represent models produced in our experiments, such as \textit{BRITS (zero imputation)}.

\subsection{Evaluation Metric} \label{eval-metric}
As mentioned before, our real-world data has data quality issues, notably, missing data and rare positive samples. Moreover, in collected positive samples, some of them are intrinsically unpredictable, such as fiber cuts. All of these factors result in our predictions having low recalls.

Considering our imbalanced datasets, we select Area Under Precision-Recall Curve (PR-AUC) as the metric, which is not sensitive to the number of samples. Formal definitions of precision and recall are displayed in Equation~\ref{precision_and_recall}. From another aspect, we weigh precision more than recall, thus targeting high precision at the expense of low recall. Therefore, we focus on the PR-AUC to recall 0.1, and consequently, the full score of our evaluation metric is 0.1. As formulated in Equation~\ref{eval_metric}, we define our evaluation metric as an integral of area under the curve, where $D$ is our evaluation metric, namely the area under PR curve to recall 0.1, axis $x$ is the recall, axis $y$ is the precision, and $f(x)$ represents the PR curve. Our evaluation metric is used for model selection and model evaluation.

\begin{equation} \label{precision_and_recall}
	\begin{aligned}
		Precision &= \frac{\text{True Positives}}{\text{True Positives} + \text{False Positives}} \\
		Recall &= \frac{\text{True Positives}}{\text{True Positives} + \text{False Negatives}}
	\end{aligned}
\end{equation}

\begin{align}
	\label{eval_metric}
	D = \int_{0}^{0.1} \mathrm{dx} \int_{0}^{f(x)} \mathrm{x y dy}
\end{align}

\subsection{Experiment Settings} \label{exp-settings}
\subsubsection{Comparison of Models on Individual Datasets}
\textbf{Missing data processing}. We apply zero-imputation and median-imputation separately on datasets for random forest. For XGBoost and BRITS, we also use zero-imputation on their inputs to make a fair comparison to random forest. Furthermore, XBoost and BRITS both have designed mechanisms to handle missing values, so we let them process missing data by themselves to show how much improvement these two mechanisms can bring.

\noindent\textbf{Hyper-parameter searching.} For tree-based algorithms random forest and XGBoost, we run parameter searching for the hyper-parameter \textit{number of trees} on both of them. The value range of \textit{number of trees} is from 100 to 3,000 in intervals of 100 and a total of 30 values. For each value, we train random forest and XGBoost on the training set. After that, we use the validation set to pick out the best model and run on the test set to generate test scores. 

In the training of BRITS, we first train it only on the imputation task until the estimation loss on the validation set tends to be steady. Next, we continue optimizing the model on both the imputation task and classification task. BRITS is trained on each network dataset by an Adam optimizer with the learning rate as $10^{-3}$ and batch size as $1,024$.

\subsubsection{Transfer Learning}
Transfer learning's first stage, namely pre-training, is applicable to all models on the mega-dataset. After the pre-training finishes, we begin to fine-tune the pre-trained model on network-specific samples. In the fine-tuning period, we lower the learning rate to $5 \times 10^{-4}$. Generally, we only allow the last few layers in the neural network to update parameters during fine-tuning. For instance, in classification tasks, only the classifier should be fine-tuned but with other parts frozen to keep the learned representation fixed and only adjust the classification border in the classifier. This is how the model \textit{BRITS (fine-tune the classifier only)} is fine-tuned. In addition, we have another fine-tuning model, \textit{BRITS (fine-tune the entirety)} that fine-tunes the entire BRITS model to observe whether adjusting the learned representation together with the classification can help obtain more improvement.

\subsubsection{Performance on Main Use-cases}
NHP has been applied on two important commercial use-cases, namely 100G OTN line cards and 10G Ethernet clients. To test our model performance on both use-cases, we extract their respective samples in the mega-test dataset and run trained models of XGBoost and BRITS to obtain test scores. Furthermore, we train XGBoost and BRITS separately on 100G OTN line cards and 10G Ethernet clients. The purpose is to compare the learning performance between models trained on single facility type and all 12 facility types.

\subsection{Experiment Results}
Our model performance results on each single-network dataset are listed in Table~\ref{tbl2}. Table~\ref{tbl3} contains the results of transfer learning across networks. Table~\ref{tbl4} lists model performance on two main use-cases in NHP: 100G OTN line cards and 10G Ethernet clients. In all tables, bold values are the best results on the dataset.

\subsubsection{Comparison Across Models Trained on Individual Datasets} \label{baseline-results}
The results in Table~\ref{tbl2} show that \textit{Random Forest (zero imputation)} and \textit{Random Forest (median imputation)} achieve comparable results on datasets Network1, Network2, and Network5. On Network3, Network4, and Network6, we observe that \textit{Random Forest (median imputation)} is inferior to \textit{Random Forest (zero imputation)} by 61.1\% of the weighted average score. The results indicate that (1) imputation methods can affect model performance significantly for random forest on our datasets; (2) the median imputation works on some of features, but not on the others since our datasets from different networks have different distributions in the feature space.

Models of XGBoost and BRITS all produce higher learning accuracy than random forest. Both \textit{XGBoost} and \textit{BRITS} deomstrate a 72.4\% performance improvement over baseline \textit{Random Forest (zero imputation)} on the weighted average score. It is observed that \textit{XGBoost (zero imputation)} produces the same learning accuracy as \textit{BRITS (zero imputation)} in terms of the weighted average score. Likewise, \textit{XGBoost} and \textit{BRITS} have the same overall performance. Furthermore, missing data handling mechanisms of XGBoost and BRITS can bring further improvement (with 4.2\% improvement on the weighted average score). Overall, \textit{XGBoost} and \textit{BRITS} have comparable learning performance on datasets Network3, Network4, and Network5. On Network2, \textit{BRITS} is 11.5\% better than \textit{XGBoost}. On Network6, \textit{XGBoost} outperforms \textit{BRITS} by 7.1\%. 

We further observe the model performance on individual networks with small datasets, in particular on Network1. Network1 has the least number of samples and the least number of positive samples, leading to least ratio of positive samples. Both models of BRITS outperform XGBoost and random forest. XGBoost's performance on this small dataset and a low ratio of positive samples is close to baseline models. This also indicates BRITS is a more suitable solution to address the challenge of low positive data ratio and the small sample size. The result of \textit{BRITS (zero imputation)} is 13.6\% better than \textit{BRITS}, indicating zero-imputation is more effective than BRITS' missing data processing mechanism on dataset Network1. Moreover, \textit{BRITS (zero imputation)} gains 38.9\% improvement than the baseline model of \textit{Random Forest (zero imputation)}. It also outperforms \textit{XGBoost (zero imputation)} by 47.1\%. As the size of data samples increases, the performances of XGBoost and BRITS become comparable.

\begin{table}[ht]
	\centering
	\caption{
		Performance comparison across models trained on single-network datasets. The assessment metric used is PR-AUC to recall 0.1, defined in Section~\ref{eval-metric}. The rightmost column shows each model's scores averaged over the mega-datasets that represent models' overall performance. The average scores are weighted by the number of samples in each test dataset. The best result of each column is highlighted in bold.}
	\resizebox{160mm}{!}{
	\begin{tabular}{p{143pt}<{\centering}|p{36pt}<{\centering}p{36pt}<{\centering}p{36pt}<{\centering}p{36pt}<{\centering}p{36pt}<{\centering}p{40pt}<{\centering}|p{60pt}<{\centering}}
		\toprule
		Model                                   & Network1    & Network2    & Network3    & Network4     & Network5     & Network6  & Weighted Avg. \\
		\midrule
		Random Forest (zero imputation)         & 0.036      & 0.010      & 0.018      & 0.023       & 0.046       & 0.042      & 0.029 \\
		Random Forest (median imputation)       & 0.035      & 0.007      & 0.003      & 0.012       & 0.049       & 0.005      & 0.018 \\
		\midrule
		XGBoost (zero imputation)                & 0.034      & 0.023      & \bf{0.087} & 0.039    & 0.066         &  0.056     & 0.048 \\
		BRITS (zero imputation)                 & \bf{0.050} & 0.026      & 0.078      & 0.039      & 0.065       & 0.053       & 0.048  \\
		XGBoost 							    & 0.033      & 0.026      & \bf{0.087} & \bf{0.041} & 0.066       & \bf{0.060}  & \bf{0.050} \\
		BRITS  							        & 0.044      & \bf{0.029} & \bf{0.087} & 0.040       & \bf{0.067}  & 0.056      & \bf{0.050} \\
		\bottomrule
	\end{tabular}
	}
	\label{tbl2}
\end{table}

\begin{table}[!htp]
	\centering
	\caption{Performance comparison across models trained on mega-datasets. The assessment metric used is PR-AUC to recall 0.1. The rightmost column shows each model's scores averaged over the mega-datasets that represent models' overall performance. The average scores are weighted by the number of samples in each test dataset. The best result of each column is highlighted in bold.}
	\resizebox{160mm}{!}{
	\begin{tabular}{p{138pt}<{\centering}|p{38pt}<{\centering}p{38pt}<{\centering}p{38pt}<{\centering}p{38pt}<{\centering}p{38pt}<{\centering}p{40pt}<{\centering}|p{60pt}<{\centering}}
		\toprule
		Model                                   & Network1    & Network2  & Network3    & Network4     & Network5     & Network6  & Weighted Avg. \\
		\midrule
		Random Forest (zero imputation)         & 0.024      & 0.009      & 0.009       & 0.022        & 0.044       & 0.028     & 0.024 \\
		\midrule
		XGBoost								    & 0.058      & 0.023      & \bf{0.089}  & \bf{0.040}   & 0.064       & 0.055     & 0.048 \\
		BRITS (pre-trained only) 				& 0.056      & 0.026      & 0.088     & 0.038       & 0.066       & 0.056      & 0.049 \\
		BRITS (fine-tune the classifier only)   & \bf{0.060} & 0.026      & 0.085      & 0.038       & \bf{0.068} & 0.056      & 0.050 \\ 
		BRITS (fine-tune the entirety) 			& 0.055     & \bf{0.030}& 0.087       & \bf{0.040}  & \bf{0.068} & \bf{0.057}  & \bf{0.052} \\ 
		\bottomrule
	\end{tabular}
	}
	\label{tbl3}
\end{table}

\begin{figure}[!ht]
	
	\subfigure[BRITS performance comparison on samples from Network1]{
		\begin{minipage}[t]{0.49\textwidth}
			\centering
			\label{sub-figure-b}
			\includegraphics[width=8cm]{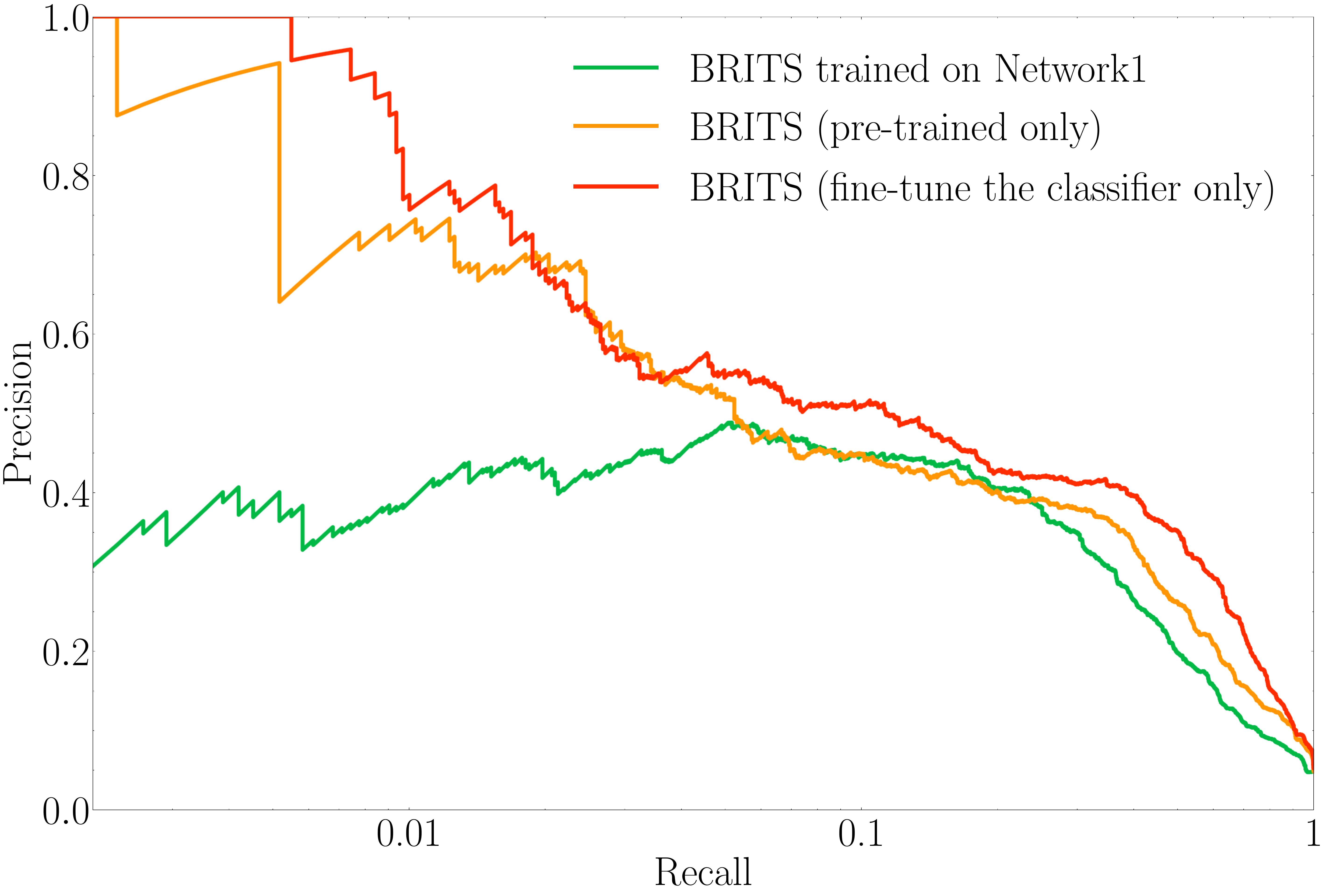}
		\end{minipage}
	}
	\subfigure[Model performance comparison on the overall mega-test set]{
		\begin{minipage}[t]{0.49\textwidth}
			\centering
			\label{sub-figure-a}
			\includegraphics[width=8cm]{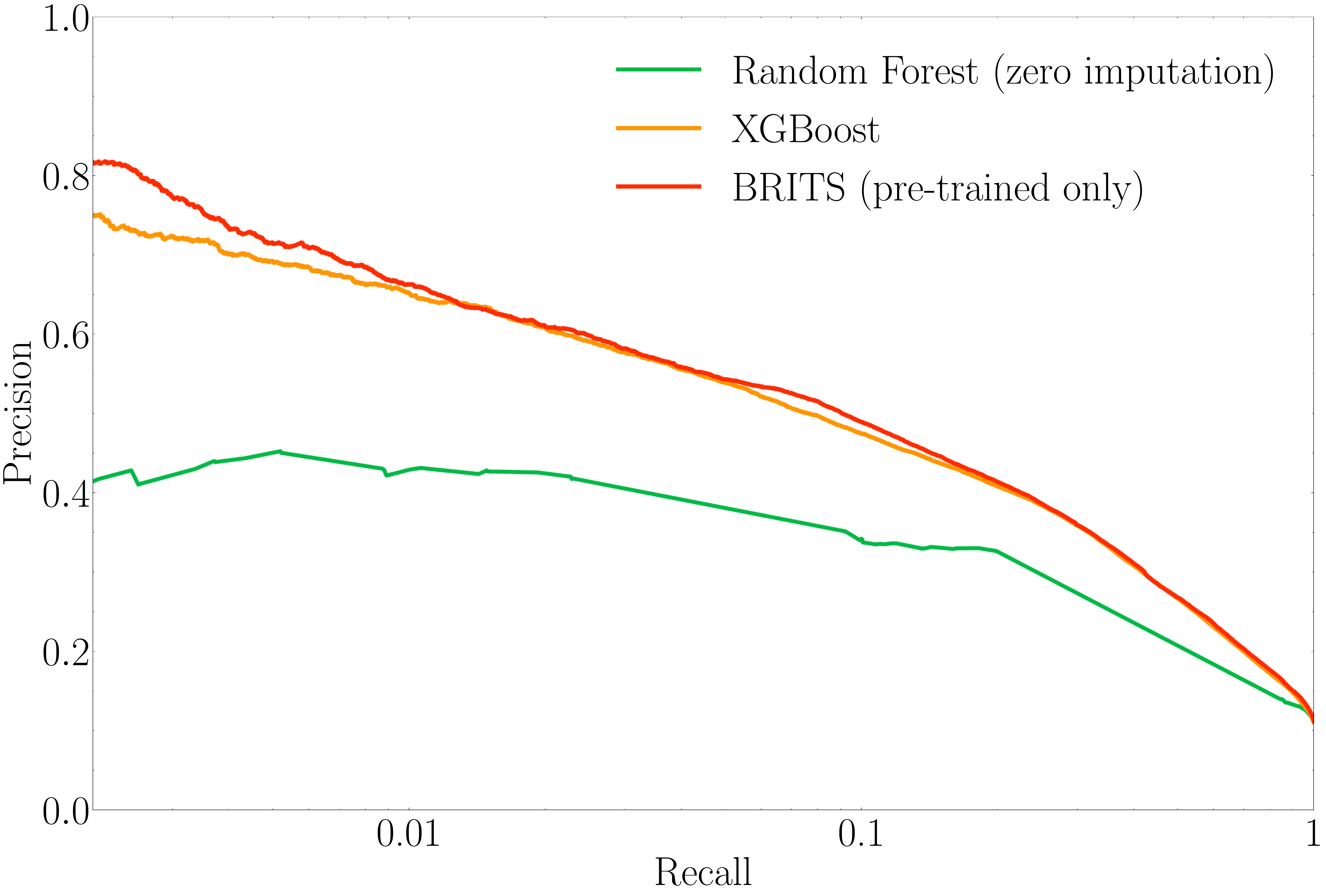}
		\end{minipage}
	}
	\caption{Performance as a function of recall for various models in this study.  We apply log scaling on the recall axis. In subfigure~\ref{sub-figure-b}, we compare the performance of BRITS models evaluated on data from Network1 only. The model is pretrained on the mega-dataset and fine-tuned on Network1.  In subfigure~\ref{sub-figure-a}, we compare the performance of \textit{Random Forest (zero imputation)}, \textit{XGBoost} and \textit{BRITS (pre-trained only)}(see Table~\ref{tbl3}), evaluated on the mega-test set. }
	\label{fig:PR_AUC_comparison}
\end{figure}

\subsubsection{Transfer Learning}
Transfer learning is adopted to improve the learning performance on networks with insufficient data samples. Transfer learning normally consists of two stages, namely pre-training and fine-tuning. Models of \textit{Random Forest (zero imputation)} and \textit{XGBoost} only have the pre-training stage without the fine-tuning stage. We can consider pre-training on mega-datasets as the knowledge sharing of six networks. As shown in Table~\ref{tbl3}, knowledge sharing through pre-training on mega-datasets allows \textit{XGBoost} to improve the learning performance on Network1 by 75.8\% compared to Table~\ref{tbl2}. The Random Forest Model does not benefit from the knowledge sharing practices with even degraded results. One possible reason is the increased missing rate in the mega-dataset. As shown in Table~\ref{tbl1}, the mega-dataset's missing rate is the highest as a result of merging 125 feature columns from six network datasets.

\textit{BRITS (pre-trained only)} is pre-trained on the mega-dataset but not fine-tuned on any network-specific samples. Meanwhile, the second variation of the transfer learning model is \textit{BRITS (fine-tune the classifier only)} that has its imputer frozen and its classifier fine-tuned on corresponding network-specific data samples. This means the representation learned from the mega-dataset is fixed while the classifier is fine-tuned. The third variation of the transfer learning model on BRITS is \textit{BRITS (fine-tune the entirety)} with both its imputer and classifier being fine-tuned.

In the case of Network1 with the smallest dataset, \textit{BRITS (pre-trained only)} achieves 27.3\% improvement in Table~\ref{tbl3} compared with \textit{BRITS} trained only singular networks in Table~\ref{tbl2}. This indicates that transfer learning from mega-dataset to singular network is effective with relatively small statistics. \textit{BRITS (fine-tune the classifier only)} further improves learning performance by 36.4\%. However, the performance of \textit{BRITS (fine-tune the entirety)} degrades compared to \textit{BRITS (pre-trained only)}. This indicates that fine-tuning for Network1 helps the classifier but not the imputer of the BRITS model. We plot Precision-Recall curves in Figure~\ref{sub-figure-b} to demonstrate the learning performance improvements. 

In contrast, on Network2/3/4/5/6 with larger datasets, all the models produce comparable learning performances in Table~\ref{tbl2} \emph{vs} Table~\ref{tbl3}. This illustrates that knowledge transfer from the mega-dataset to a singular network stops improving the models' accuracy with sufficiently large statistics. However, the models' performance does not degrade either, which indicates that a single pre-training from the mega-dataset can produce ML models usable for all networks at once. 

To demonstrate the model performance on the overall mega-dataset, Precision-Recall curves plotted in Figure~\ref{sub-figure-a} display a comparison between \textit{Random Forest (zero imputation)}, \textit{XGBoost}, and \textit{BRITS (pre-trained only)}. \textit{XGBoost} and \textit{BRITS (pre-trained only)} both outperform \textit{Random Forest (zero imputation)}. In comparison, \textit{BRITS (pre-trained only)} has comparable learning performance to \textit{XGBoost}. 

\subsubsection{Performance on Main Use-cases}
\begin{table}[!htp]
	\centering
	\caption{Comparison of model performance on two important use-cases. Models presented in this table are all trained on the mega-dataset, but on different subsets of the 12 facilities. For example, \textit{XGBoost trained on 100G OTN line cards} is trained on samples collected from only line-facing ports of 100G OTN line cards. The evaluation metric used is PR-AUC to recall 0.1. Models trained on 12 facility types obtain close performance with models specifically trained on 100G lines or ETH10G clients.}
	\begin{tabular}{c|cc}
		\toprule
		Model                                 & 100G OTN line cards  & 10G Ethernet clients \\
		\midrule
		XGBoost	trained on 100G OTN line cards               & 0.049 & /    \\
		XGBoost	trained on 10G Ethernet clients              & /     & 0.047    \\
		XGBoost trained on all 12 facility types             & 0.049 & \bf{0.053}   \\
		\midrule
		BRITS trained on 100G OTN line cards                 & 0.054 & /       \\
		BRITS trained on 10G Ethernet clients                & /     & 0.050    \\
		BRITS trained on all 12 facility types               & \bf{0.055} & 0.052    \\
		\bottomrule
	\end{tabular}
	\label{tbl4}
\end{table}

To validate XGBoost and BRITS' performance on two main commercial use-cases, we compare models trained on all 12 facility types with models trained on only 100G lines or ETH10G clients. As the results show in Table~\ref{tbl4}, for both XGBoost or BRITS, models trained on 12 facility types produce comparable results with models specifically trained on 100G lines or ETH10G clients. Model \textit{XGBoost trained on all 12 facility types} in Table \ref{tbl4} is equivalent to \textit{XGBoost} in Table \ref{tbl3}; and \textit{BRITS trained on all 12 facility types} in Table \ref{tbl4} is equivalent to \textit{BRITS (pre-trained only)} in Table~\ref{tbl3}. XGBoost and BRITS achieve comparable learning performance on the use case of ETH10G clients. On the use case of 100G lines, BRITS models perform approximately 10\% better than XGBoost models.

\begin{figure}[!ht]
	\centering
	\includegraphics[width=10cm]{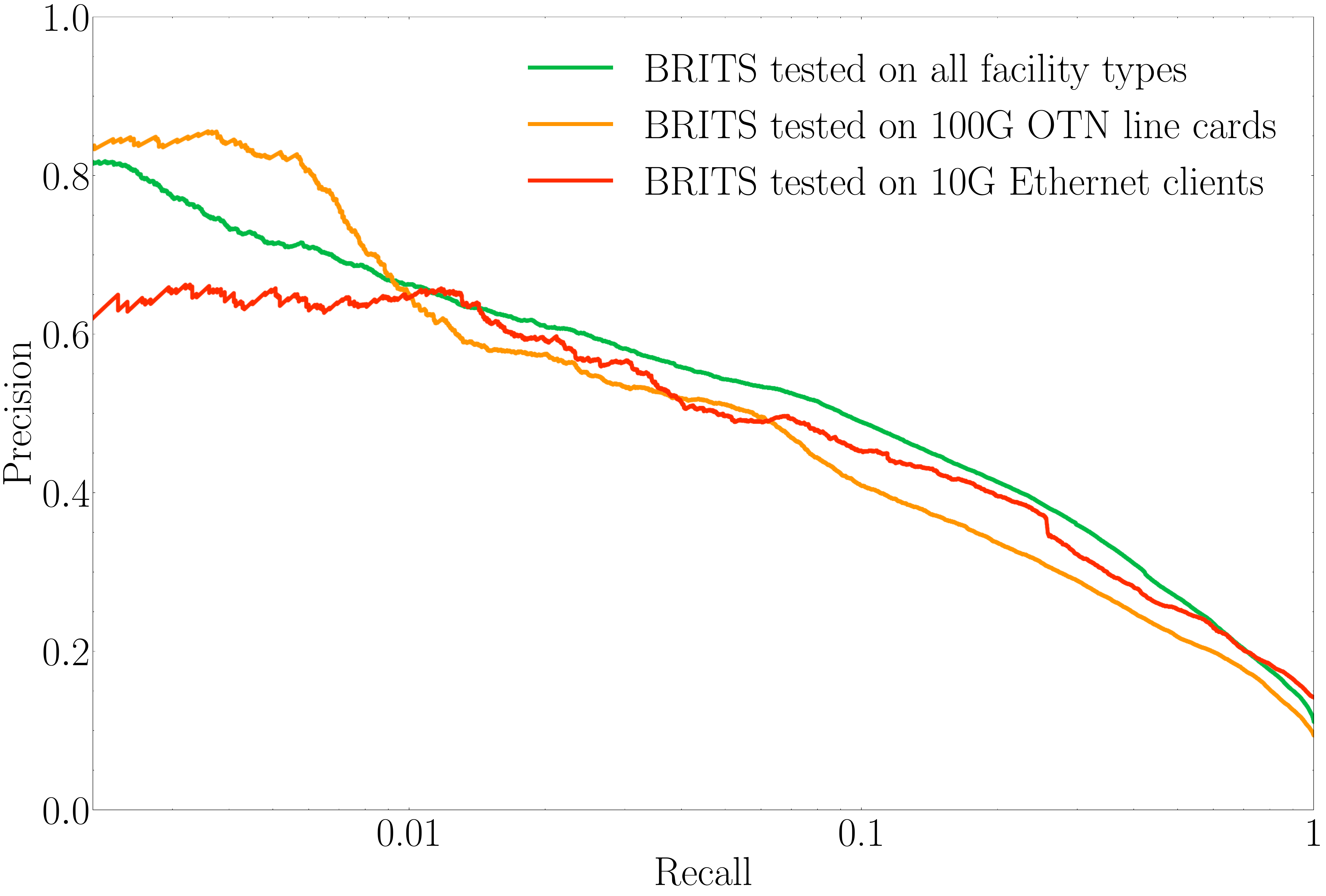}
	\caption{Precision as a function of recall for the model \textit{BRITS (pre-trained only)} in Table~\ref{tbl3} evaluated on on all facility types (green) the 100G line-side facilities (orange) and 10G Ethernet client-side facilities (red).}
	\label{fig:PR_curve2}
\end{figure}

Figure~\ref{fig:PR_curve2} contains PR curves of \textit{BRITS (pre-trained only)} tested separately on samples of all 12 facility types, 100G lines, and ETH10G clients. Over all facility types, our model achieves 1\% recall at 65\% precision. For 100G lines, the precision can go above 80\% at low recall. As a result, 3\% of network outages are reported at least once 1-7 days before they occur. In practical large-scale networks, this corresponds to roughly a dozen alerts per day. Given the limited resources that network operators have to investigate forecasted failures, as opposed to managing current actual failures, this is a practicable number. Operators benefit from receiving a relatively small number of forecasted alerts, which have a high probability of being correct, rather than being flooded with a large number of alerts, which would require many resources to investigate.

\subsubsection{Summary}
Our rule-based labeling method encodes the domain knowledge to identify ILOS events automatically.  The generated datasets with ILOS labels enable our supervised ML methodology for forecasting. XGBoost is an out-of-box model on the ILOS prediction task with the benefit of easy and fast training. As a deep learning model, BRITS obtains similar results to XGBoost and performs better on small datasets. With transfer learning, BRITS obtains the best weighted average score. Furthermore, knowledge sharing further improves XGBoost's performance on small datasets as well. To address the missing data problem, zero-imputation works relatively well on our datasets because zero-suppression is the dominating factor for missing data in counter PMs. However, we observe that more sophisticated missing-data handling in XGBoost and BRITS brings additional improvements and produces the optimal learning performance. Last but not least, the singular model of XGBoost or BRITS trained on mega-datasets can handle all 12 facility types from all six networks without the need of training the model for each facility type or each network specifically.

\section{Conclusion} \label{sec-conclusion}
This paper has developed a supervised ML methodology to forecast Imminent Loss of Signal (ILOS) in the long term, 1-7 days before they occur in optical networks.  To achieve this goal, we have developed ML best practices tailored for real-world telecommunications data. We have analyzed PM data collected from optical network equipment in six commercial networks of different sizes, totalling 1,373 days and 16,456,331 data samples. As expected, we discover that the general robustness of optical networks and the prevalence of unpredictable events such as fiber cuts in the field make LOS events relatively rare and unpredictable. Still, our models are able to forecast 3\% of ILOS events from all receiver ports of the optical network with 65\% precision per prediction. For 100G OTN lines, our precision reaches over 80\%, albeit at low recall. Furthermore, we achieve nearly optimal results with a single ML model for all facilities and all networks, making it usable for commercial products that work out-of-the-box. Hence this work enables preventive actions by network operators to avoid network outages.

The suppression of zero values and the diversity of metrics reported by ports on these networks result in extremely sparse input datasets to our ML models, of which 70\%$\sim$80\% are null values. To overcome this obstacle, we have experimented with XGBoost and BRITS ML architectures, which both can handle null input values. With all missing values imputed with 0, we find that our XGBoost and BRITS models both outperform the baseline random forest model significantly. Furthermore, letting XGBoost and BRITS handle missing data by themselves can further obtain a small improvement. Without transfer learning, BRITS obtains similar results to XGBoost. While XGBoost is easy to use and fast to train, the ability of the BRITS architecture to impute missing values enables other application scenarios, such as generating complete data for other tasks. With the help of transfer learning, \textit{BRITS (fine-tune the entirety)} achieves the best results on our datasets. For network datasets with insufficient data to train a robust ML model, knowledge sharing across all six networks can improve the model performance significantly (\textit{XGBoost} improved by 75.8\%, \textit{BRITS (fine-tune the classifier only)} improved by 36.4\%). Fine-tuning on BRITS deliver improvement, but not as much as knowledge sharing that shares general representation of ILOS signatures across networks.

In future work, we plan to improve our ML methodology by considering multiple prediction classes corresponding to different root-causes and failure modes in optical networks. This will enable better separation of intrinsically unpredictable events from predictable ILOS and better optimization of the ML training for the latter. We will also leverage data with higher time-resolution and loosen the strict distinction between future-tense forecasting vs present-tense assurance, which will increase the recall and multiply the added value of the application for network operation centers. Finally, we plan to extend our analysis to multi-vendor equipment, as we expect our methodology to apply qualitatively to any optical network from which similar performance monitoring data can be obtained.

\section*{Acknowledgments}
Thanks to Tom Chen and Jincheng Sun for their early technique work. Thanks to Benjamin Freund, Emil Janulewicz, Minming Ni, Patrick Prémont, Thomas Triplet, Sothy Phan, Shelley Bhalla and Bill Kaufmann from the Blue Planet software team. Thanks to Paul Gosse, Dana Dennis and Yinqing Pei for expert guidance about Ciena equipment. Thanks to Mitacs for facilitating a fruitful collaboration between Concordia University and Ciena.

{
	\bibliographystyle{unsrt}
	\bibliography{references}
}

\end{document}